\documentclass[aip,graphicx, reprint,11pt]{revtex4-1}

\usepackage{graphicx}
\usepackage{subfigure}
\usepackage{hyperref}
\usepackage{cleveref}
\usepackage{xcolor}

\begin{document}


\title{Understanding the surface wave characteristics using 2D particle-in-cell simulation and deep neural network} 

\author{Rinku Mishra}
\email{rinku.mishra@ipr.res.in}
\affiliation{ Centre of Plasma Physics, Institute for Plasma Research, Nazirakhat, Sonapur 782402, Assam, India.}
\affiliation{Institute for Plasma Research, HBNI, Bhat, Gandhinagar 382428, India}

\author{S. Adhikari}
\affiliation{%
Department of Physics, University of Oslo, PO Box 1048 Blindern, NO-0316 Oslo, Norway
}%
\author{Rupak Mukherjee}
\affiliation{%
Princeton Plasma Physics Laboratory, Princeton, New Jersey, 08540, USA
}%

\author{B. J. Saikia}
\affiliation{ Centre of Plasma Physics, Institute for Plasma Research, Nazirakhat, Sonapur 782402, Assam, India.}

\begin{abstract}
The characteristics of the surface waves along the interface between a plasma and a dielectric material have been investigated using kinetic Particle-In-Cell (PIC) simulations. A microwave source of GHz frequency has been used to trigger the surface wave in the system. The outcome indicates that the surface wave gets excited along the interface of plasma and the dielectric tube and appears as light and dark patterns in the electric field profiles. The dependency of radiation pressure on the dielectric permittivity and supplied input frequency has been investigated. Further, we assessed the capabilities of neural networks to predict the radiation pressure for a given system. The proposed Deep Neural Network model is aimed at developing accurate and efficient data-driven plasma surface wave devices. 
\end{abstract}


\maketitle 

\section{\textbf{Introduction}}
Waves in the bounded plasma have gained considerable interest in the last few decades. While studying the interaction of an electromagnetic field with the plasma, along with bulk wave modes, it is also important to consider the wave mode concentrated at the surface.Such modes are known as surface waves. Such waves play a very crucial role in understanding plasma turbulence in the divertor of laboratory fusion devices\cite{lee2021collisional}, torus of fluid\cite{novkoski2021experimental}, plasma diagnostics, laser physics\cite{luo2013laser}, communication, atomic spectroscopy\cite{hubert1996atomic}, high-density plasma generations, and even in plasma processing. Furthermore, in astrophysical plasma, surface wave energy transport has an important role in understanding the magnetosphere and solar corona problems \cite{gordon1983collisional,shu1992volume,roth2010surface,buti1990solar}. Steinolfson et al.\cite{steinolfson1986viscous} found that at higher frequencies, the viscous damping of surface waves can cause the heating of the solar corona. Even in quantum plasma, surface wave study has made quite good imprints. To name a few, Lazar et al.\cite{lazar2007surface} investigated the dispersion properties of the surface wave by incorporating the quantum statistical pressure as well as the quantum tunneling effects and observed that these effects make it easier for the electrostatic surface waves to propagate in plasma. Shahmansouri et al.\cite{shahmansouri2017exchange}, investigated the influences of Coulomb interaction along with other quantum forces and the external magnetic field to understand the properties of ion-acoustic surface waves. 

The surface wave is an electromagnetic wave localized at the surface, which propagates along with the interface of mediums having different permittivity. The name comes from the notion of carrying energy, mainly in the near interface region, and decays exponentially away from the interface. Such nature of a wave is also known as an evanescent wave. In plasma physics, the surface wave propagation takes place only if the dielectric permittivity of the medium is negative \cite{briggs:1970}. For cold plasma approximation, the permittivity is described as $\epsilon_p = 1 - \omega_{pe}^2/\omega^2$, where $\omega_{pe}$ is the plasma frequency and $\omega$ is the wave frequency. When such a medium with negative permittivity is surrounded by a dielectric of positive permittivity, $\epsilon_d$, the propagation takes place along the interface with frequency $\omega$ $\leq$ $\omega_{pe}/(1+\epsilon_d)^{1/2},$ where $\epsilon_d $ is the relative permittivity of glass\cite{moisan1975small}. Upon satisfaction of the given condition, plasma can sustain the surface wave with an evanescent field on both the sides\cite{landau1960}. Such nature of the wave aids in dumping the energy into the system, resulting in rapid ionization in the plasma medium.
Plasma surface waves were first observed experimentally by Trivelpiece and Gould \cite{trivelpiece1959space} using a cylindrical plasma column bounded by dielectric. After that, the surface wave has been investigated considering the interface separating the media of two different dielectrics\cite{janaki1998surface,Mishra_2018,cramer1996alfven,ivanova1993radiophysics, Mishra_void}.  For many early years, high-frequency discharges have been maintained between the metal electrodes or the resonant cavities. However, the method was very complex to maintain the discharge. Then a new way of plasma production was proposed that can reduce some complexity of already existing devices based on RF and microwave frequencies. The new method uses electromagnetic (EM) surface waves to sustain the discharge\cite{moisan1975small}. Eventually, the interest in such waves attracted researchers to use in sustaining plasma columns. Therefore, in the 1970s, a simple, compact, and well-organized device named surfatron was developed, which uses electromagnetic surface waves to generate plasma columns at microwave frequencies\cite{moisan1975small,moisan1991plasma}. It does not require any external magnetic field along the column (plasma) axis or other additional devices to sustain the plasma, like other devices. As the electrodes are not required for the wave excitation, the problem of gas contamination and the corrosion of the electrodes due to interaction with the plasma gets reduced. Apart from the ease of plasma sustainability, such systems have amazing applications in the field of communication technology as plasma antennas. A plasma antenna is the best alternative for a traditional metal antenna with re-configurable input capability. Several articles reported the theoretical and experimental investigation of the surface wave and its characteristics\cite{kaw1970,margot1993,Baruthram1993,lee1995,lee2000} in the context of its usage as antenna. However, there are still open questions regarding surface wave physics yet to be answered that would provide a new perspective about the field.\cite{ott:2008,omelchenko:2012}
 
Besides the laboratory and industrial investigation, several authors have studied surface waves using different numerical simulation techniques. To begin with, Kousaka et al.\cite{kousaka2002numerical} studied the propagation of electromagnetic waves along the plasma dielectric interface wave using the 2D finite difference time domain (FDTD) approximation. Igarashi et al.\cite{igarashi2004finite} investigated the plasma surface wave using the finite element method. Kabouzi et al.\cite{kabouzi2007modeling} used a self-consistent two-dimensional fluid-plasma model coupled with the Maxwell equations for surface wave sustained argon plasma discharge.  Despite all these investigations described above, we believe that there is a significant gap in understanding regarding the kinetic properties of plasma in support of the surface wave propagation.

In the present work, we have investigated surface waves and their kinetic characteristics. There are various factors on which the surface wave propagation depends. Parameters like plasma density, source frequency, and surface material permittivity are a few of the important ones among those factors. Out of which, we have considered the source frequency and the material permittivity to study the surface wave characteristics. To quantify the impact of the aforementioned parameters, we have observed the radiation pressure pattern in the presence of surface waves. The work is performed using particle-in-cell (PIC) simulation \cite{birdsall1991plasma} of an argon plasma with XOOPIC\cite{verboncoeur1995object}. The code uses the Monte Carlo collision algorithm  \cite{vahedi1995monte} to model collisions with neutrals in the system. The result presented in this work is believed to add new physical aspects of the surface wave sustained plasma column to aid future experimental investigations and innovations.

One of the unique aspects of the present work comes in the form of application of Deep Neural Network (DNN) to predict the radiation characteristics of a given system. Deep Learning is a sub-field of machine learning in Artificial intelligence (A.I.) that deals with algorithms inspired by biological neurons. It uses sophisticated mathematical modeling to process data in complex ways. DNN consists of neural networks that have an input layer, an output layer, and at least one hidden layer in between. In the neural network, the input layer is the first layer that accepts the external information or data and passes it into the input layer's units. The output layer is the last layer of the neural network that produces outputs for the model. It provides the outcome or prediction of the data fed into the input layer. Hidden layers are the intermediate layer between the input and output layers. There can be either one or more hidden layers in a network. Being a combination of neural network and machine learning, DNN provides a perfect tool to leverage deep learning for all machine learning tasks and expect better performance with surplus data availability. With the advent of cost-effective computing power and data storage, deep learning has been embraced in every digital corner of our everyday life. However, there is a big gap in research-based physical models as of now. The usability of DNN is unlimited if a user can train such a model with physics-based parameters. In recent studies, it has shown promising outcomes in terms of accurate prediction of physical quantities\cite{cheng2021deep,raissi2019physics,lagaris1998artificial,adhikari:2021}. 
The need for such a model appears due to computationally expensive kinetic codes such as PIC. For example, a typical run of XOOPIC with the presented system configuration takes 10-12 hours (wall-time) depending on the number of particles used. In order to reduce statistical noise, that number can ever go up to 36-40 hours. Now, this is why a consolidated artificially trained model is essential such that prediction of radiation data becomes numerically cheap. In the present work, we have built a DNN model comprising a complex neural network of several layers to predict estimated radiation data for a given plasma system. Additionally, we have made it open-source, such that people can contribute their data to improve the accuracy of their model and experimental facilities. \\
The paper has been organized as follows. Section II presents the simulation model for the study, followed by results and discussion of the work in section III. Section IV describes the implementation and usage of the Deep Neural Network (DNN) in the present work. Lastly, we conclude the work in section V.

\section{Simulation method and modeling}
In the present work, the interaction of plasma with an electromagnetic field inside a dielectric tube has been studied using a Particle-in-Cell simulation code (XOOPIC\cite{verboncoeur1995object}). It is an X-Windows version of OOPIC, which is a two-dimensional relativistic electromagnetic object-oriented particle-in-cell code written in the C++ programming language. . The PIC method is preferred over others to better understand the non-linear properties of plasma and kinetic behavior. It models a plasma system consisting of a large number of superparticles (having the same charge to mass ratio as the real plasma particles) distributed in a spatial domain. The charge and current densities are calculated using the superparticle's position and velocity data in a spatial domain. Maxwell's equations are solved at each time step for discrete mesh points, and particles are moved by the resulting forces calculated on the particles. A detailed description of the algorithm of XOOPIC code can be found in a paper written by Verboncoeur et al. \cite{verboncoeur1995object}. It has features of modeling two spatial dimensions in both Cartesian (x,y) and cylindrical(r,z) geometry, including all three velocity components. The applicability of this code ranges from plasma discharges, such as glow and RF discharges, to microwave-beam devices. The code can handle an arbitrary number of species and includes Monte Carlo Collision (MCC) algorithms for modeling intra-species collisions of charged particles as well as collisions with the neutral background gas. 

\begin{figure}[ht]
     \includegraphics[width=\linewidth]{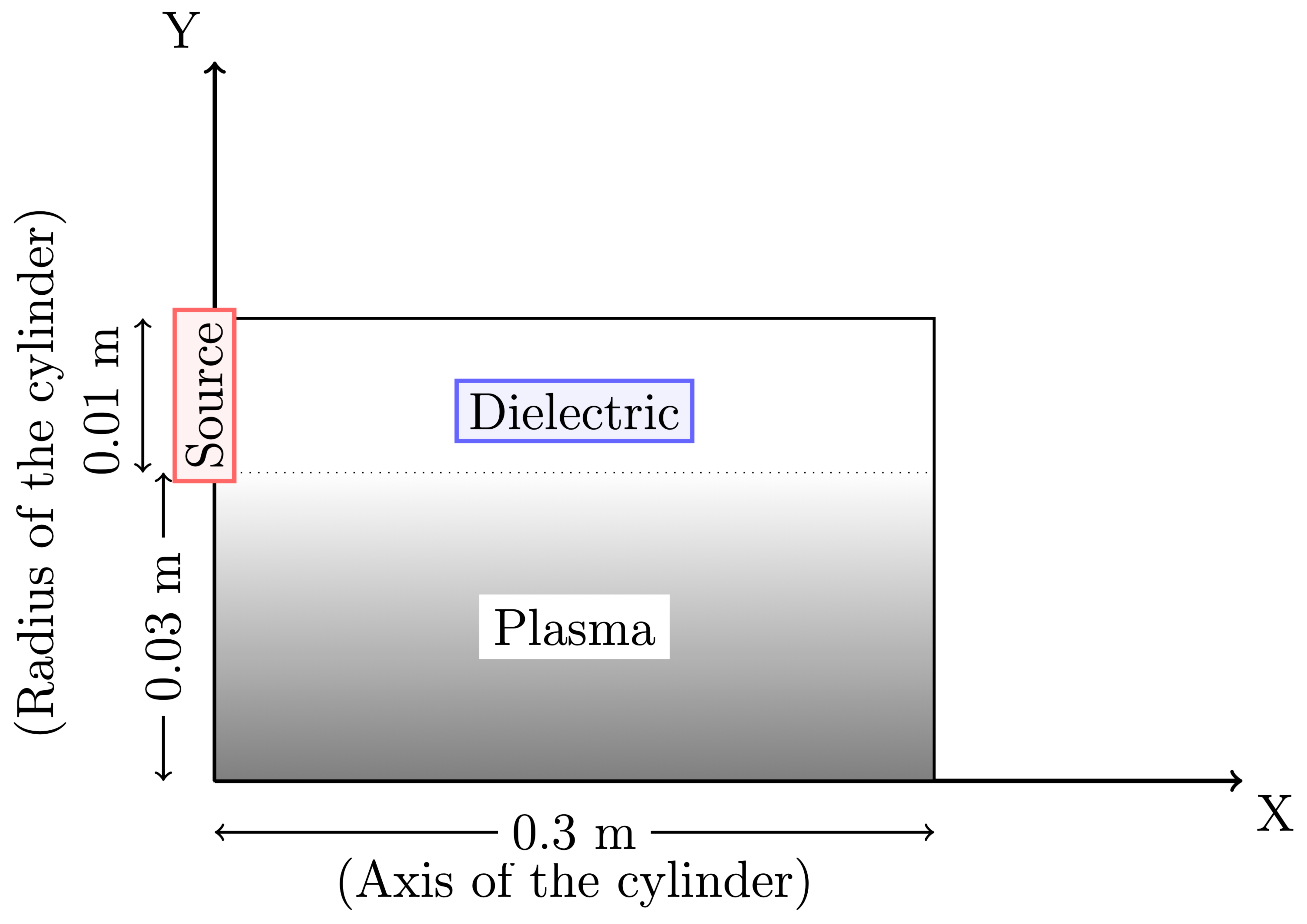}%
     \caption{ Schematic of our simulation model. Length of the tube in x, and y direction are 0.3 m and 0.03 m respectively. The thickness of the dielectric tube is 0.01 m. Source of microwave frequency 2.45 GHz is placed at the left top of portion of the model. The left and right boundary of the system is considered to be absorbing. }
     \label{fig:1}
\end{figure}

The model is developed considering the symmetric half of the cylinder on Cartesian geometry. The schematic of our model is given in \cref{fig:1}. The length of the cylinder is considered to be $0.3$ m in the x-direction and $0.03$ m in the y-direction. The top boundary is a dielectric material with a thickness $0.01 $ m. The left and right boundaries of the plasma system are considered to be nearly perfect conductors (for the fields) and to absorb the incident particles. Particles (electrons and Argon ions) are loaded uniformly throughout the domain from Maxwellian distributions with respective temperatures $T_e$ and $T_i$ (see \cref{table:1}). The plasma density is sustained by ionization via collision with the constant background neutrals (pressure of $0.1 $ Torr). However, the ionization of the second and higher-order argon atoms is neglected. A source of microwave frequency ($f_{source}$) at $2.45 $ GHz is added using a wave port (at the left top portion of the model). The dielectric constant ($\epsilon_{r}$) of the surrounding material and the source frequency ($f_{source}$) are continuously varied to quantify the effects on radiation profile. For the baseline simulation, we have set $f_{source}$ at $2.45$ GHz and $\epsilon_{r}$ at $5$.

To maintain the stability of the numerical method, step size and grid size has been chosen carefully, satisfying the Courant condition \cite{de2013courant}. The spatial grid is composed of 2048 ($Nx \times Ny$) cells of cell size $0.00468 \times 0.00125$ m, which ensures that Debye length is resolved. On average, $39$ computational particles (super-particle) per cell are loaded, resulting in adequate resolution for the field quantities with minimum statistical noise. As mentioned, a time step of $10^{-12}~s$ is chosen to meet Courant criteria, ensuring electrons (fastest particle) never cross an entire spatial cell in one single time step. For each case, the simulation was run for $2\times10^6$ timesteps, which translates into $2~\mu s$. In our modeling approach, we load a uniform density at the beginning in a similar fashion as others \cite{cooperberg1998series,Matsumoto:M96/87} and run the simulation long enough to make sure the loss of particles at the walls and generation of particles due to charge-neutral collision is balanced. Each case was run to reach a quasi-stationary state and terminated when there were no significant changes in the plasma density profiles.

As we mentioned in the introduction, a typical run of XOOPIC can take 10-12 hours (wall-time) or may be even 36-40 hours based on grid resolution and statistical weight. We made a moderate choice of 20 hours per run for 1024 parametric variations. It approximately took 42 days using HPC (parallel and sequential).

\begin{table*}
\begin{ruledtabular}
\begin{tabular}{llr}
    Parameter & Value\footnote{The values have been adapted from the work of N. Matsumoto\cite{Matsumoto:M96/87}.}\\
    \hline
    Initial Plasma density ($N_{e,i}$) &  $ \sim 10^{14}~m^{-3}$  \\
    Final Plasma density ($N_{e,i}$) &  $\sim 1.5 \times 10^{16}~m^{-3}$ (baseline)  \\
    Electron temperature ($T_e$) &  2~eV \\ 
    Ion temperature ($T_i$)  &  0.03~eV   \\ Debye length ($\lambda_d$) & $0.001$ m\\
    Number of  cells ($N_x$, $N_y$) & 64, 32  \\ 
    Time step ($\delta t$) & $10^{-12}$ s \\
    Spatial Grid size ($\delta x$, $\delta y$) & $0.00468$ m, $0.00125$ m \\
    System length ($L_x$, $L_y$) & 0.3 m, 0.04 m\\
    No. of particles  &  102400 (baseline) \\ 
    Neutral Gas  & Argon (Ar)\\
    Pressure & 0.1 Torr \\
\end{tabular}
\end{ruledtabular}
\label{table:1}
\caption{Simulation Parameters.}
\end{table*}

\section{\label{sec:3} Results and discussions}
\begin{figure*}[ht]
     \subfigure[]{\includegraphics[width=0.45\linewidth]{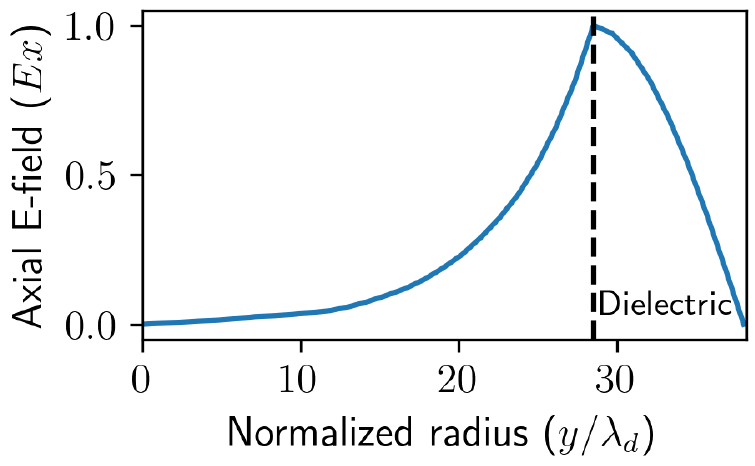}}
     \subfigure[]{\includegraphics[width=0.45\linewidth]{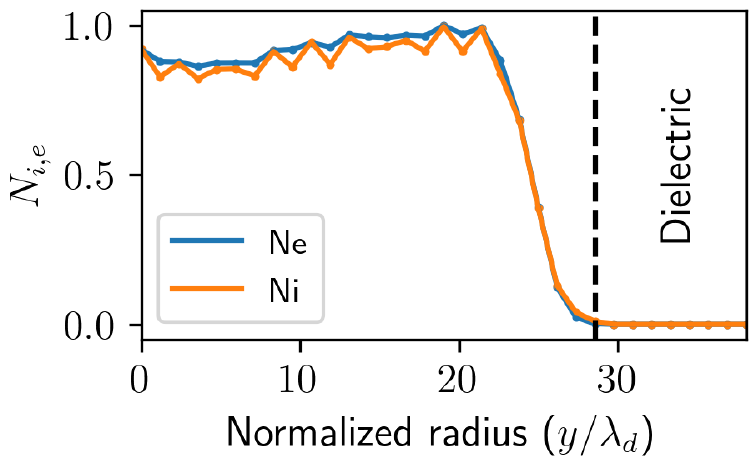}}
     \caption{(a) Axial electric field and (b) density profiles along the radial direction in across the plasma and the dielectric medium. The black dotted line represents the interface of the plasma and the dielectric material. The frequency and the dielectric constants are considered to be $2.45$ GHz and 5 respectively. The axial electric field normalized by its maximum value (maximum electric field ~$1.5\times10^4$ V/m) and the density is normalized by the bulk plasma density $1.5\times10^{16}~m^{-3}$. The radial length is normalized by the electron Debye length ($\lambda_d$) for both figures. 
    }
     \label{fig:2}
\end{figure*}

The physical properties of the surface wave have been investigated using the parameters given in \cref{table:1}. Under the application of a microwave source, the interaction of the electromagnetic field with the plasma leads to the localization of surface charge near the interface region (charge bunching). The presence of surface charge causes the maximum electric field intensity in the region and the intensity decreases as we move away from the interface region (in the radial direction). Such behavior is known as an evanescent nature. Existing literature on plasma surface wave indicates that the evanescent nature is one of the assurances for the surface wave to exist. \Cref{fig:2} describes the spatial profile of the normalized electric field ($E_x$) and the densities along the radial direction (y-direction) for the baseline simulation. Here, the electric field has been normalized with its maximum value (maximum electric field ~$1.5\times10^4$ V/m) and the system radius (y) with electron Debye length. For the densities, bulk plasma density is taken as reference for normalization. The frequency and the dielectric constants for this figure have been considered $2.45$ GHz and $5$, respectively. It can be observed from the \cref{fig:2} that near the dielectric surface, the field intensity is maximum and then decays in the radial direction, which points toward the existence of a surface wave. The general idea of surface wave study is based on the epitome of reality that only surface involved is the interface between two mediums. The surrounding vessel are the artifact of a more practical approach. The general approach of surface wave study is to neglect the effect of the plasma sheath. The argument for such consideration is the sheath scale length, $\lambda_{De}$, which is much less than the typical scale length (depth of penetration) of the surface wave\cite{cooperberg1998surface,cooperberg1998series,moisan1982properties,lee2007kinetic,lee2010kinetic}. However, one can find such dispersion including the effect of sheath in the work by Cooperberg et al. \cite{cooperberg1998electron,cooperberg1998nonuniform}. For our case, we have also observed the existence of sheath near the interface region as shown in \cref{fig:2}(b).

\begin{figure*}[ht]
  \centering
  \subfigure[]{\includegraphics[width=0.45\linewidth]{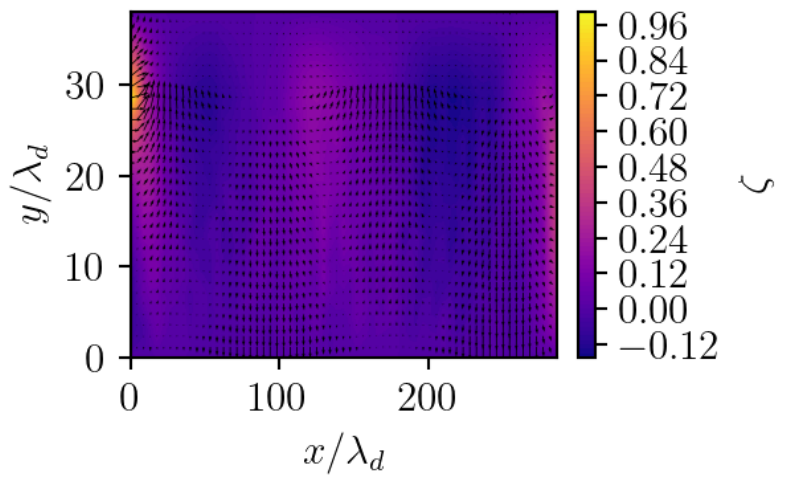}}
  \subfigure[]{\includegraphics[width=0.45\linewidth]{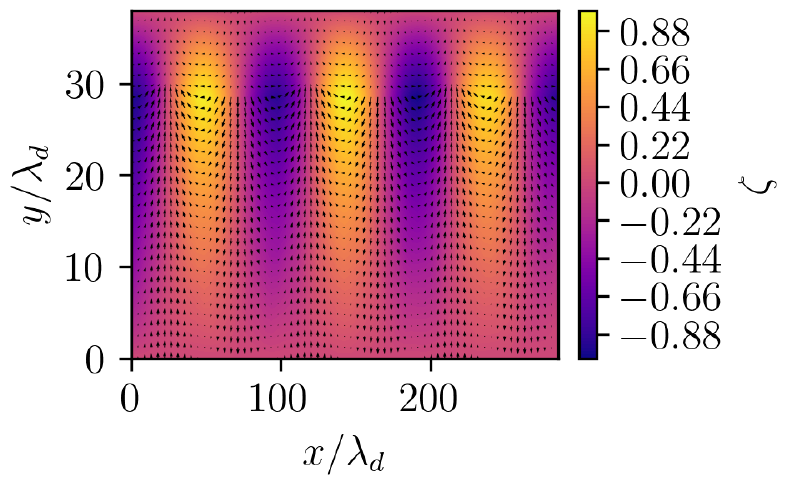}}
  \subfigure[]{\includegraphics[width=0.45\linewidth]{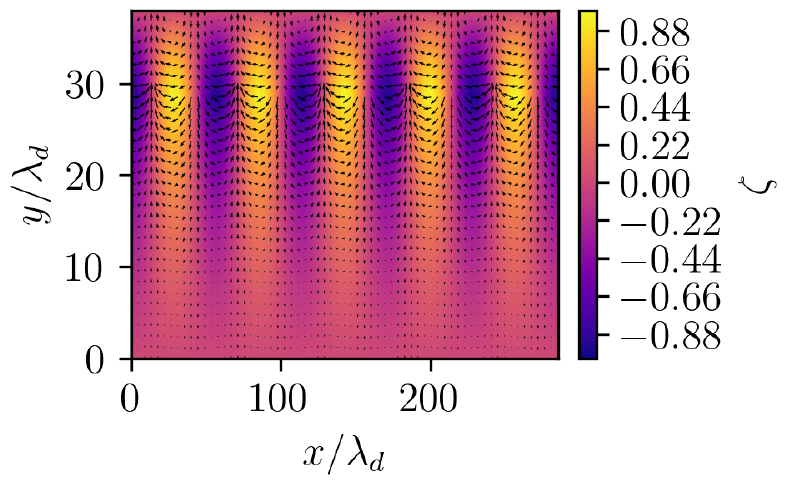}}
  \subfigure[]{\includegraphics[width=0.45\linewidth]{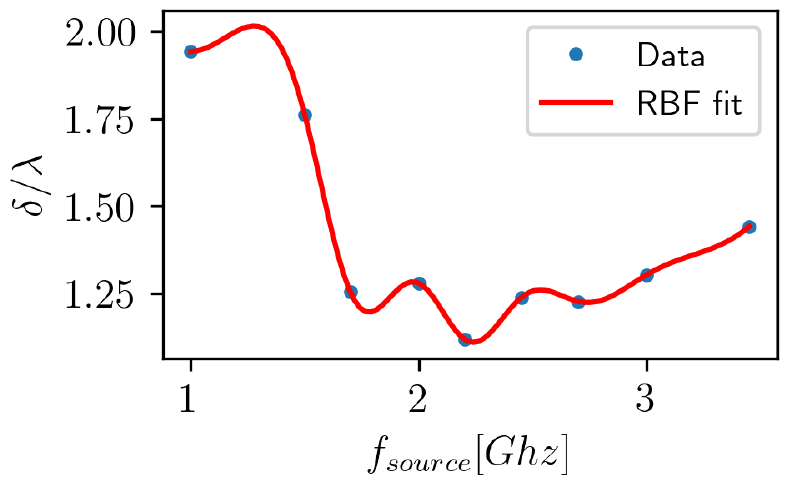}}
\caption{ Profile of axial electric field intensity near the interface region for different frequency values. (a)Frequency - $1.7$ GHz (maximum electric field intensity ~$3.6\times10^3~V/m$), (b)Frequency - $2.45$ GHz (maximum electric field intensity ~$6.4\times10^3~V/m$), (c) Frequency - $3.45$ GHz (maximum electric field intensity ~$1\times10^5~V/m$). The colorbar in the figures represents the ratio of electric field intensity to its maximum value denoted by $\zeta$, (d) Pattern of the charge bunching width depending on the different frequency values using radial fit. $\delta$ in the figure represents the difference of the space charge width. The normalization of this parameter has been done by the wavelength of the wave $\lambda$.}
\label{fig:3}
\end{figure*}

The basic idea of surface wave generation in such systems is based on superposition of oppositely directed traveling waves from two different regions near the interface.  Such superposition produces standing wave in which the plasma oscillates with certain frequency known as surface wave frequency\cite{trivelpiece1959space}. In \cref{fig:3}, we observe the normalized axial profile of the electric field, $E_x(x,y)$, normalized with its maximum value. The presence of positive and negative charge separated regions confirms the existence of standing waves in that region (shown as bright and dark patches in \cref{fig:3}). The plasma particles get trapped and oscillate with surface wave frequency. There is various literature also which supports our theory of surface wave\cite{zhelyazkov1978stimulated,trivelpiece1959space,cooperberg1998series}. The color bar in the figures represents the intensity of the electric field denoted by $\zeta$.

The other observation from this field intensity profile is, at a frequency below the $2.45$ GHz frequency, the present model does not observe any charge bunching near the interface region (shown in \cref{fig:3}(a)). The reason might be the low energy input source, which might not be strong enough to increase the plasma density in order to excite the surface wave. The increasing frequency enhances the collision between the plasma particles and neutrals, exceeding the critical density. It causes the bunching of the surface charge near the interface leading to the excitation of surface wave propagation. However, for further increase in input frequency, we have also observed a decrease in the charge bunching width. Using the Radial Basis Function (RBF), fit for the data , \cref{fig:3}(d), shows the trend of charge-width variation depending on the frequency value. RBF method has been used for fitting as this method provides an excellent interpolant for high dimensional data sets of poorly distributed data points. In the figure $\delta$ represents the difference of the charge width. The normalization of this parameter has been done by the wavelength of the injected wave, $\lambda$.

\begin{figure*}[ht]
  \centering
  \subfigure[]{\includegraphics[width=0.45\linewidth]{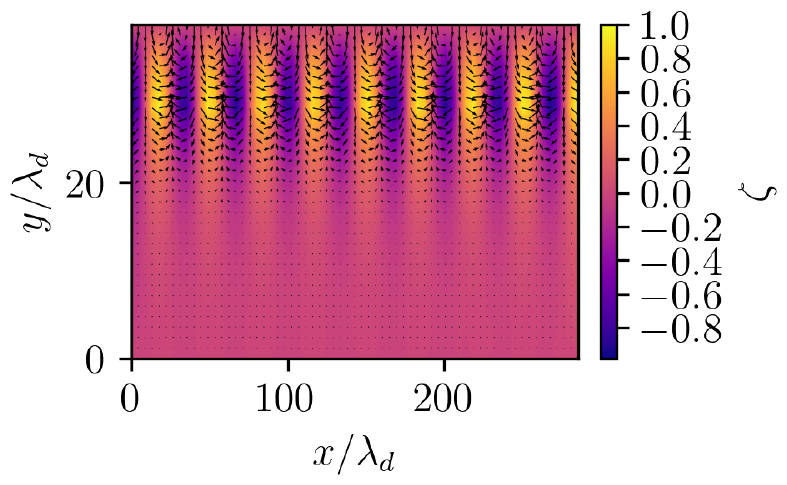}}
  \subfigure[]{\includegraphics[width=0.45\linewidth]{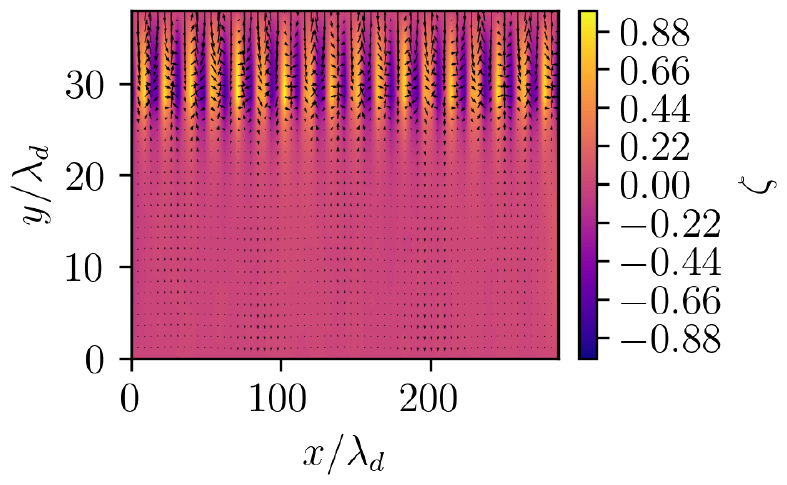}}
  \subfigure[]{\includegraphics[width=0.45\linewidth]{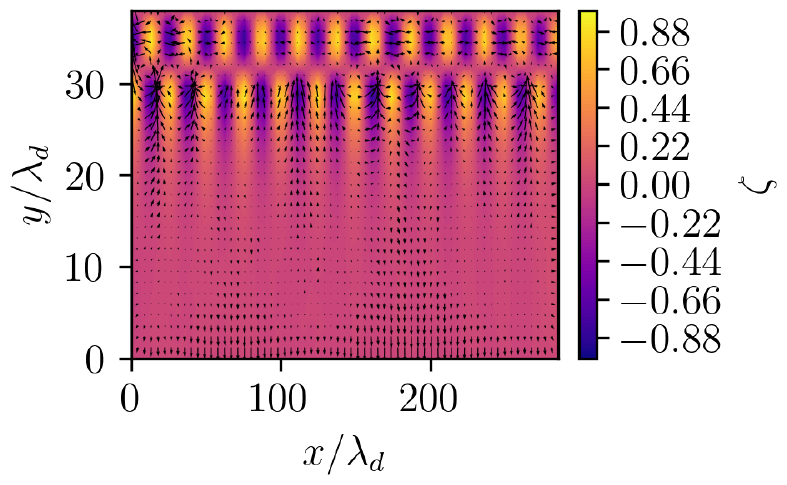}}
  \subfigure[]{\includegraphics[width=0.45\linewidth]{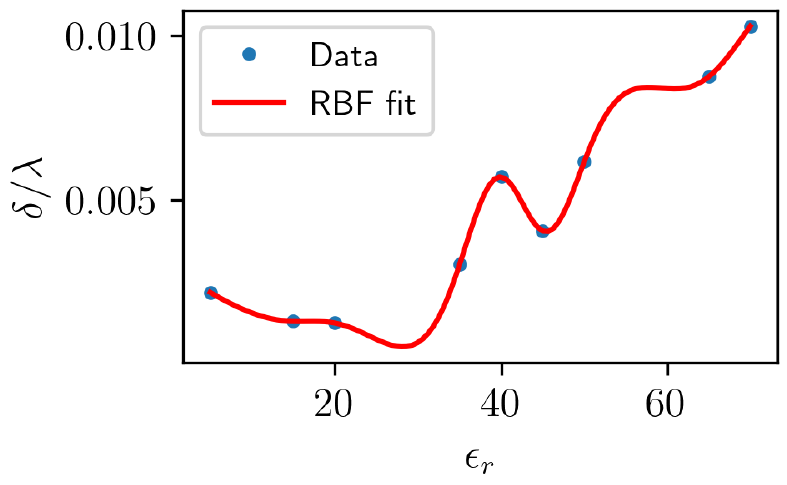}}
\caption{ Profile of axial electric field intensity near the interface region for different permittivity variations.(a) permittivity - 20 (maximum electric field intensity ~$1.6\times10^4~V/m$), (b) permittivity - 50 (maximum electric field intensity ~$7.2\times10^3~V/m$), (c) permittivity - 100.(maximum electric field intensity ~$1\times10^4~V/m$). The colorbar in the figures represents the ratio of electric field intensity to its maximum value denoted by $\zeta$. (d) Pattern of the charge bunching width depending on the different permittivity values using radial fit. $\delta$ in the figure represents the difference of the charge width. The normalization of this parameter has been done by the wavelength of the wave $\lambda$.}
\label{fig:4}
\end{figure*}

Typically, glass is used as dielectric material to study surface wave propagation in plasma experiments. However, we perform a much generalized and detailed numerical study by considering the different permittivity values. The dependency of field intensity on the surrounding permittivity materials has been shown in \cref{fig:4}. Increasing material permittivity increases the conductivity, and the charge separation width decreases. Also the wave energy is concentrated mainly near the interface region. Moreover, increasing the dielectric value of the material makes the surface behave very differently, eventually the wave patterns get destroyed (shown in \cref{fig:4}(c)). In \cref{fig:4}(d), the trend of charge bunching width variation is shown depending on the permittivity value using the RBF fit. 

\begin{figure*}
    \subfigure[]{\includegraphics[width=0.6\linewidth]{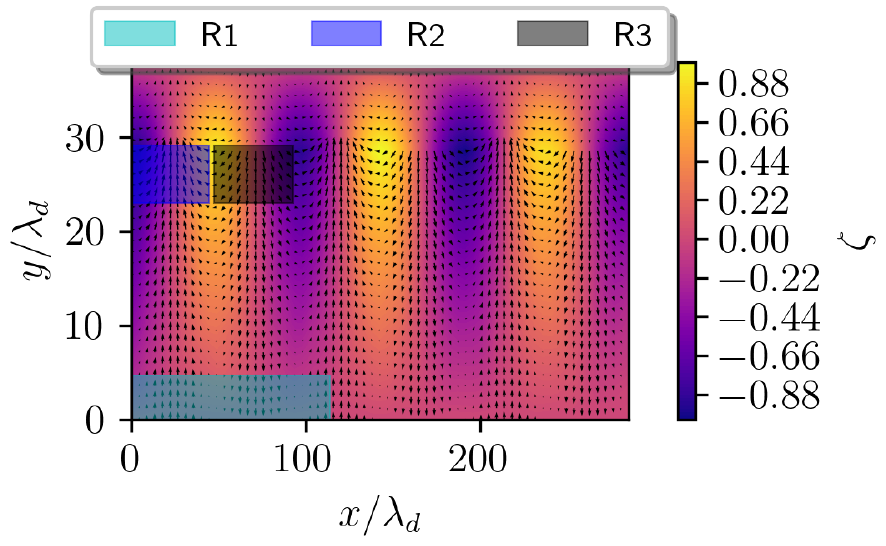}}
    \subfigure[]{\includegraphics[width=0.8\linewidth]{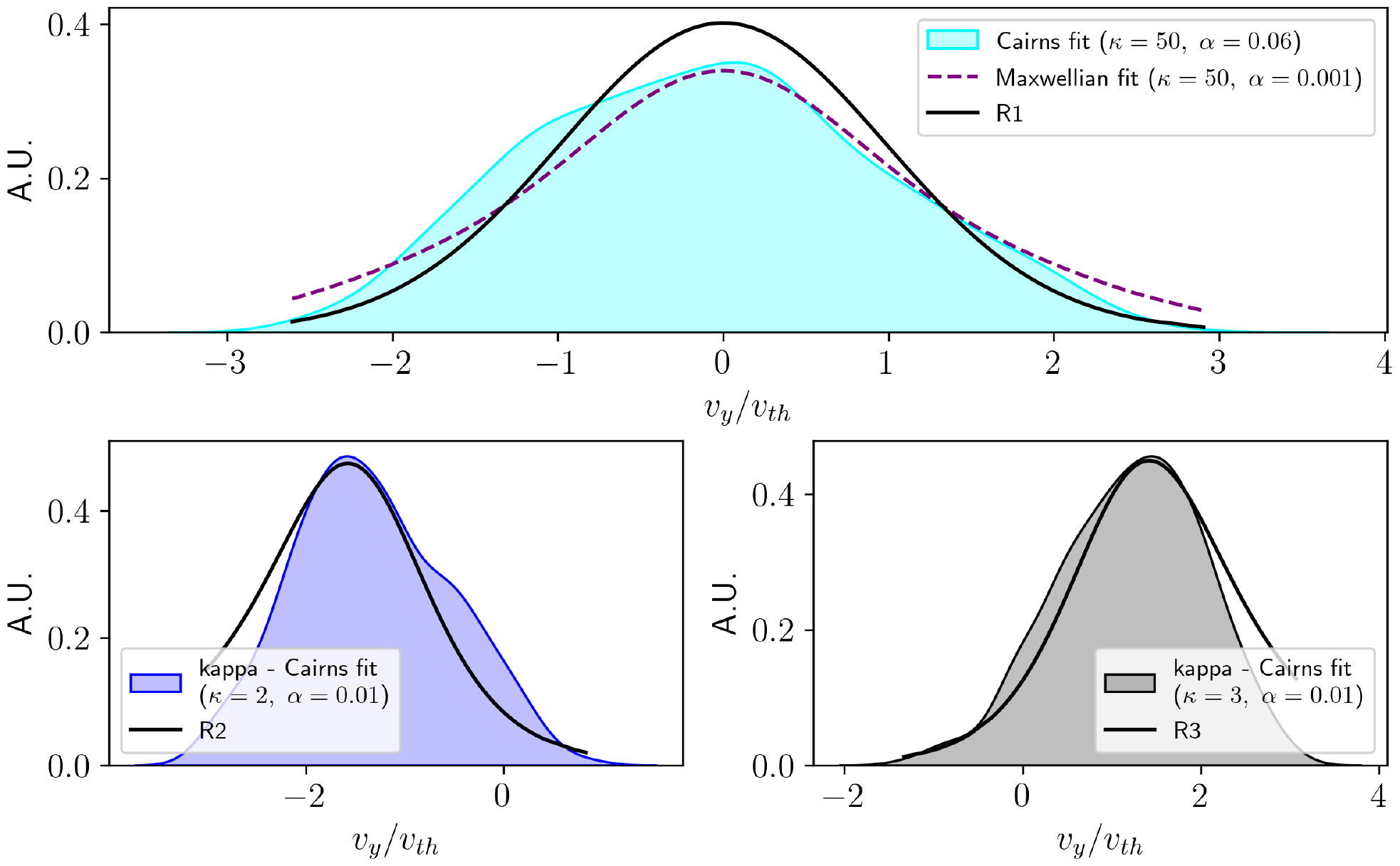}}
     \caption{(a) Axial electric field intensity with highlighted areas used for velocity reconstruction. (b) Radial velocity distributions ($f(v_y)$) of the electrons for different regions. The cyan profile represents the particle distribution in bulk region of the plasma i.e., R1 (0.0 - 114.2), The blue profile represents the particle distribution with left polarity, R2 (0.0 - 44.7) and black represents the distribution with right polarity, R3 (47.5 - 88.5). A.U is the arbitrary unit and $v_{th}$ is the thermal velocity of electrons. The bracketed numbers corresponding to each region are presented in normalized length scale ($\lambda_d$).} 
     \label{fig:5}
\end{figure*}
The velocity distribution of the electrons has been studied to understand their energy dynamics throughout the system (\cref{fig:5}b). The reconstruction of the velocity distribution is performed from the phase space at three different regions (see \cref{fig:5}a). Region R1 (0.0 - 114.1 $\lambda_d$) represents the bulk region of the system away from the interface. Regions R2 (0.0 - 44.7 $\lambda_d$) and R3 (47.5 - 88.5 $\lambda_d$) correspond to the different charge bunching regions with opposite polarity near the interface. Since we are interested in localized sampling, we also took a range in the radial direction for each axial range. For region R1, the radial range was taken as (0.0 - 4.7 $\lambda_d$), whereas for regions R2 and R3 (23.0 - 28.5 $\lambda_d$). To have a one-to-one correspondence, one should refer to the yellow and blue region in the electric field intensity profile (\cref{fig:3}b) for regions R2 and R3. Yellow and blue represent the localized positive and negative space charge regions, respectively.  

To understand the distribution of particles in the presence of reverse-polarity structure, Cairns et al.\cite{cairns1995electrostatic} introduced a particular non-thermal distribution known as Cairn's distribution. Due to the evanescent nature of the surface wave field the magnitude of the field is higher near the interface region. Therefore, the particles gains more energy in that region compared to the bulk. The effect can be observed in the velocity distribution of the electrons in region 2 and region 3. The distribution obtained here is non-Maxwellian showing the high energy tail. Such deviation from Maxwellian distribution mostly occurs when the collisions between the high-energy particles and neutral particles are infrequent compared to the low energetic particles. The underlying reason for such occurrence is the mean free path of the higher energetic particles, which is proportional to $v^4$, where $v$ is the velocity of the particles\cite{bara2014combined,aman2020revisiting,hadi2019kinetic}. The quadratic factor comes from the Fokker-Planck theory, where the slowing downtime, $\tau_s  \sim v^3$. The mean free path ($\lambda_{mfp}$) can be expressed as a product of particle velocity and slowing downtime, $v\tau_s$. Therefore, the mean free path bears the quadratic term. The detailed derivation is available in R. S. Marshall, and P. M. Bellan's work \cite{marshall2019acceleration} in addition to textbooks\cite{krallbook,bellan2008fundamentals}. Particles having such mean free path does not relax to a Maxwellian distribution state and shows the deviated distribution. Hence, the best fit for the profile naturally suits a Kappa-Cairns distribution. Such Kappa-Cairns distribution is usually found due to the presence of highly energetic non-thermal electrons and the co-existence of both the positive and negative potential. This might explain the reason behind the particles with the shifted Kappa-Cairns distribution near the interface region. The different polarity of the field defines the cause for the acceleration of the particles in different directions.

The Kappa-Cairns distribution:
\begin{widetext}
\begin{equation}
f_{e}\left(\mathbf{v} ; v_{\mathrm{th}}, \kappa, \alpha\right)= A_{\kappa, \alpha}\left(1+\alpha \frac{v^{4}}{v_{\mathrm{th}}^{4}}\right) \times\left(1+\frac{v^{2}}{2\left(\kappa-\frac{D}{2}\right) v_{\mathrm{th}}^{2}}\right)^{-(\kappa+1)},
\end{equation}

where,
$$
A_{\kappa, \alpha}=\frac{\Gamma(\kappa+1) / \Gamma\left(\kappa+\frac{2-D}{2}\right)}{\left(2 \pi v_{\mathrm{th}}^{2}\left(\kappa-\frac{D}{2}\right)\right)^{\frac{D}{2}}\left[1+D(D+2) \alpha \frac{\kappa-\frac{D}{2}}{\kappa-\frac{D+2}{2}}\right]}, ~\kappa>\frac{D}{2}+1,
$$

$v$ is the electron velocity (simulation data), $D=1,2,3$ for 1D, 2D, and 3D systems, $\kappa$ and $\alpha$ are the spectral indices of the distribution function, $v_{th}=\sqrt{\frac{kT_e}{m_e}}$ is the thermal velocity for the electrons.

For shifted Kappa-Cairns distribution, 

\begin{equation}
f_{e}^{shift}\left(\mathbf{v} ; v_{\mathrm{th}}, \kappa, \alpha\right)= A_{\kappa, \alpha}\left(1+\alpha \frac{v^{4}}{v_{\mathrm{th}}^{4}}\right) \times\left(1+\frac{(v-v_{shift})^{2}}{2\left(\kappa-\frac{D}{2}\right) v_{\mathrm{th}}^{2}}\right)^{-(\kappa+1)},
\end{equation}
\end{widetext}

In region 1, further away from the surface, the electron distribution starts to take the form of a Maxwellian. The slight deformity in the distribution might be due to the high energetic particles (produced through the interaction of plasma with electromagnetic wave) near the interface can possibly move towards the bulk region. The bulk plasma is quite close to the interface (0.03 m away, as shown in the model).  We believe this could be one of the reasons for obtaining the Cairns distribution or shifted Maxwellian distribution in the bulk region. If the system size in the radial direction is taken large, particles may lose their energy interacting with the background. In such case we may end up with a perfect Maxwellian distribution. However, it is needless to say such small change does not alter the dynamics of the system. Therefore, we did not modify the system radius as it would increase the computational hours without significant improvement of the results. Although, we could not confirm the reason behind the resemblance of the distribution towards Cairns fit in the present context. The above finding might be the scope for future works.

\begin{figure}[ht!]
  \subfigure[]{\includegraphics[width=\linewidth]{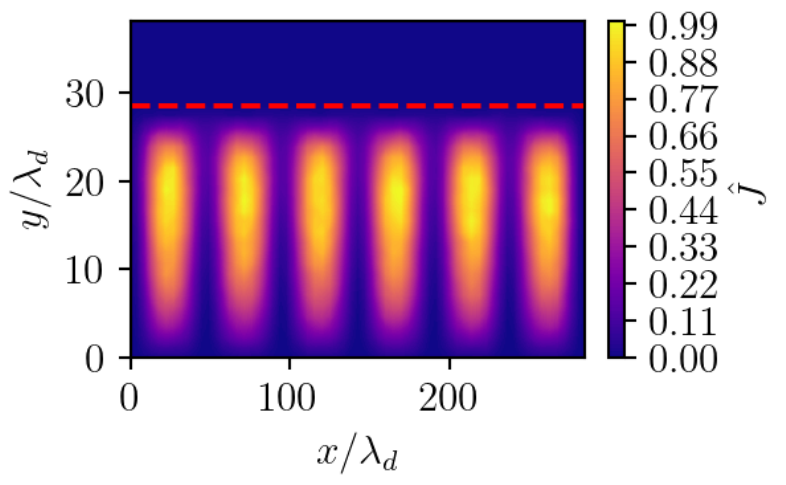}}
  \subfigure[]{\includegraphics[=\linewidth]{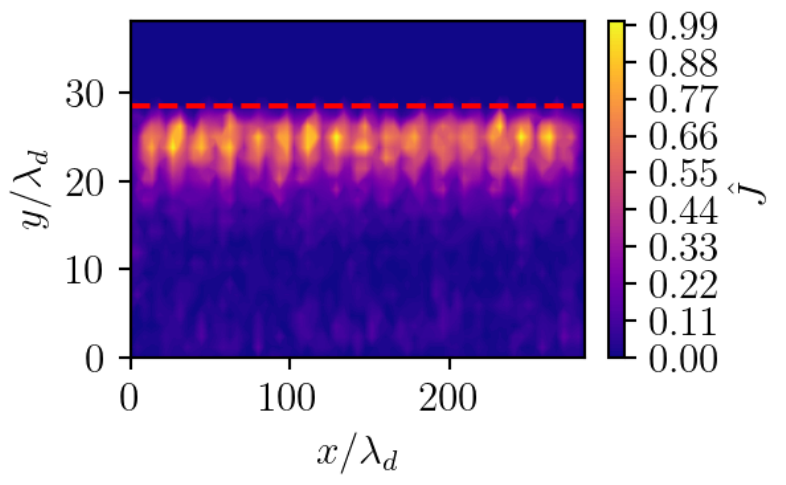}}
\caption{Spatial profile of the power deposited in the plasma by the field. The red dashed line represents the interface separating the plasma and the dielectric. The value of the power density is represented by $J$ and is normalized by its maximum value and denoted by $\hat{J}$. (a) Source frequency - $2.45$ GHz,  and permittivity - 5 (maximum value of power deposited is $~ 10^4 ~\mathrm{Wm}^{-2}$), (b) Source frequency - $2.45$ GHz  and permittivity - 20, (maximum value of power deposited is $\sim 10^3 ~\mathrm{Wm}^{-2}$).}
\label{fig:6}
\end{figure}

In \cref{fig:6}, the spatial profile of the time-averaged power density is deposited into the electrons that causes the heating of the plasma near the interface. In the earlier investigation, it has been found that at high-pressure range, ohmic heating is the dominant heating mechanism in RF plasma system\cite{cooperberg1998series}. Such a heating mechanism is responsible for the visible glow near the interface region. The appearance of such glow occurs due to the short mean free path for inelastic scattering, which causes the fast-moving electrons to lose energy quickly while moving away from the interface. The profile of electron heating shows the wavelike structure due to the presence of strong resonant surface wave fields. This makes the non-uniform heating profile of plasma in x-direction (axial direction). The strong field causes the generation of hot electron population in that region. Permittivity values in figures (a) and (b) have been used as 5 and 20, respectively, and the source frequency is considered to be $2.45$ GHz. The value of the power density is represented by $J$ and is normalized by its maximum value and denoted by $\hat{J}$.

\begin{figure}[ht!]
  \subfigure[]{\includegraphics[width=\linewidth]{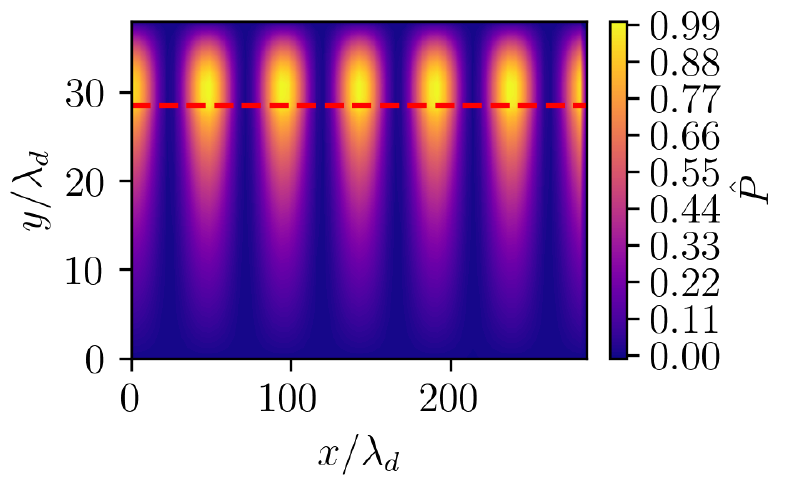}}
  \subfigure[]{\includegraphics[width=\linewidth]{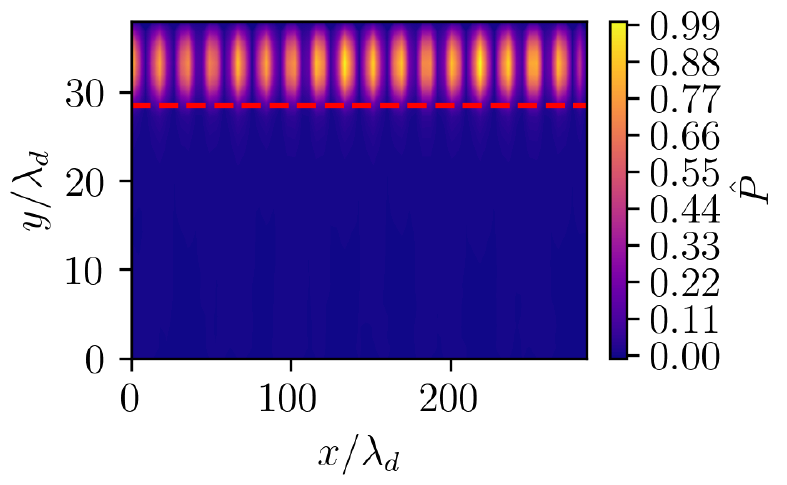}}
\caption{Radiation pressure in y-direction. The red dashed line represents the interface separating the plasma and the dielectric. The value of the radiation pressure is represented by P and is normalized by its maximum value and denoted by $\hat{P}$. (a) radiation profile for source frequency $2.45$ GHz and permittivity value 5 (maximum radiation pressure $ ~10^{-4}~\mathrm{Nm}^{-2}$), (b) radiation profile for source frequency $2.45$ GHz and permittivity value 20 (maximum radiation pressure $~10^{-5}~\mathrm{Nm}^{-2}$).}
\label{fig:7}
\end{figure}
The radiation profile of the system is given in \cref{fig:7}. The presence of localized electric and magnetic fields near the interface region shows strong peaks in that region. As the radiation pressure represents the energy loss in the medium, the value of relative permittivity decides how the radiation will behave in the system. In \cref{fig:7}(a), it can be observed that for low permittivity (for the baseline parameters), the waves in the plasma system loose their energy as they move towards the bulk region (diminishing radiation pattern). However, for higher permittivity ($\epsilon_r$ = 20 and $f_{source}$ = $2.45$ GHz) the dielectric material contains the injected wave and absorbs all the emitted radiation into the dielectric medium (shown in \cref{fig:7}(b)). We have observed the same trend for even higher permittivity values. The take away is there is a threshold for permittivity to be used as a surrounding material for surface wave instruments.

\section{Deep Neural Network Model}

In this section, we have built a Deep Neural Network (DNN) model using Particle-in-Cell simulation data sets to predict the estimated radiation data based on the system parameters of any given system. 

\begin{figure}[ht]
{\includegraphics[width=\linewidth]{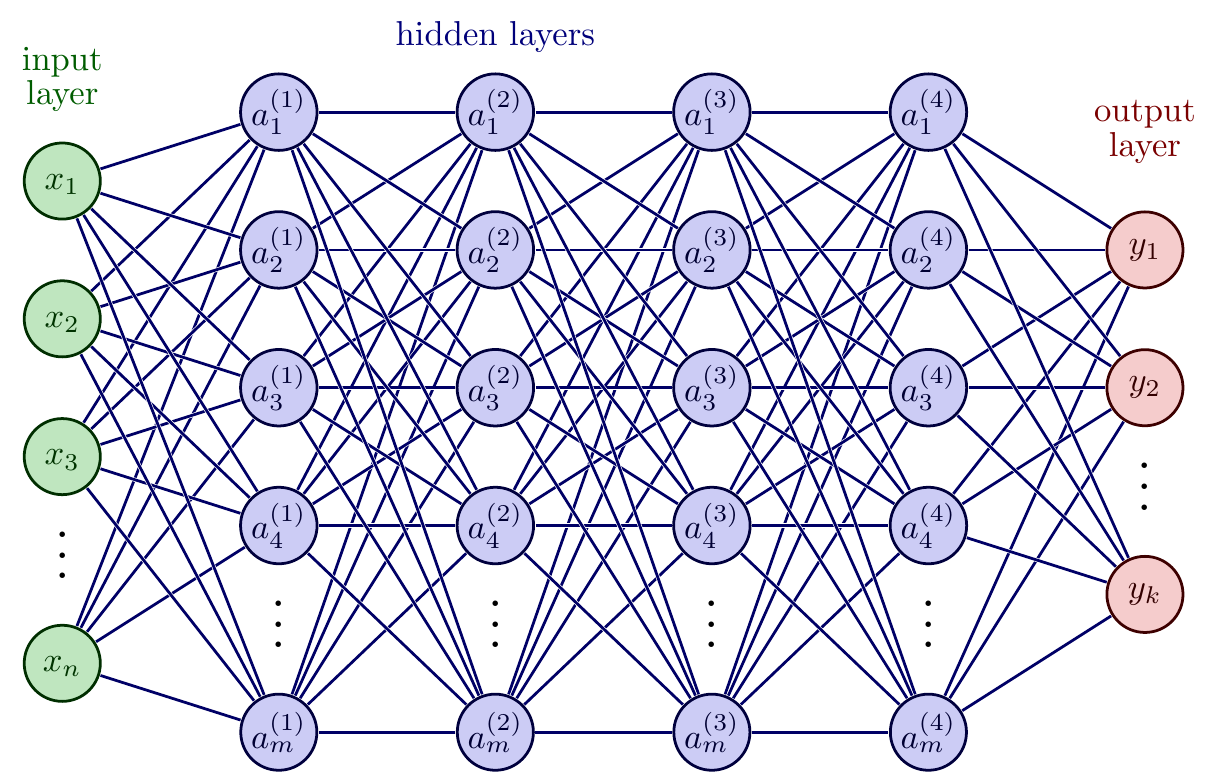}}
\caption{Schematic of the DNN model. A derivative illustration of ``Izaak Neutelings, \href{https://tikz.net/neural_networks/}{TikZ.net}; licensed under a \href{https://creativecommons.org/licenses/by-sa/4.0/}{Creative Commons Attribution-ShareAlike 4.0 International License}."}
\label{fig:8}
\end{figure}
\begin{figure}[ht]
{\includegraphics[width=\linewidth]{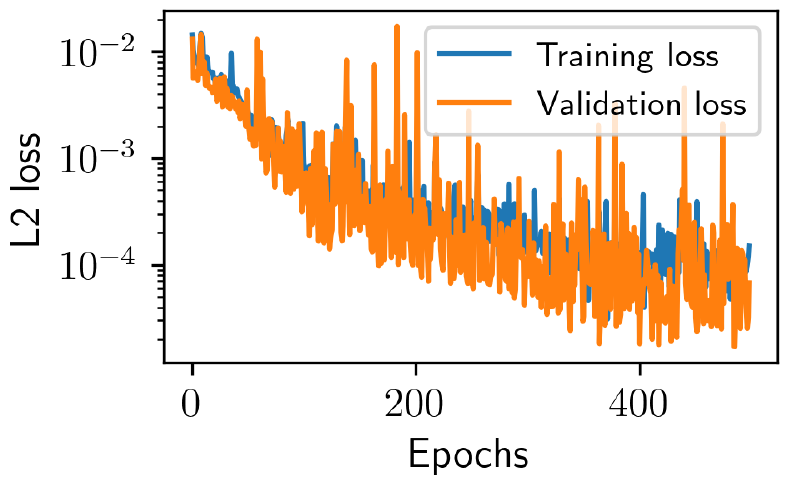}}
\caption{Training and validation loss as a function of the number of epochs for the DNN model.}
\label{fig:10}
\end{figure}

Figure \cref{fig:8} represents the schematic of a standard DNN model for such a purpose. The input layer ($x_{i}$, $i=0,1,2,...,n$, where $i$ = no. of dependent parameters or degrees of freedom) has been constructed from critical system parameters like input source frequency, dielectric permittivity of the surface material. The hidden layers ($a_{l}^{(j)}$, where, $j=1,2,..,4$, and $l=1,2,..,m$) have been developed using standard \href{https://www.tensorflow.org/}{TensorFlow} models from Keras (e.g. relu, selu, elu etc.). The number of hidden layers has been kept limited to 4 for the present work. The output layer ($y_{o}$, where, $o=1,2,..,k$) will give us the spatial dependence of radiation pattern for a specific system input. For the present case, we have considered $k=1$. After several tests, we finalized our multilayer perceptrons (MLPs\cite{gardner1998artificial,rumelhart1985learning}) with four hidden layers each consisting 100 neurons (nodes). \cref{fig:9} has been introduced to visualize neural networks' architecture and understand how the different layers are connected with different neurons. For the MLP, we have used two specific element-wise nonlinear functions called layer activation functions from Keras\cite{kerasactivation}, Rectified Linear Unit (RELU), and Scaled Exponential Linear Unit (SELU). A variant of the stochastic gradient descent algorithm referred to as ADAptive Moment estimation or ADAM has been used for the training runs. One of the important reasons for training an MLP is to determine the weight and biases using training data for the activation functions. The weights and biases depend on the choice of the loss function. For the present model, we have opted the standard choice for regression problems, the Mean Squared Error (MSE). In \cref{fig:10} the training and validation loss for the model is plotted as a function of the number of epochs. We have used 25\% of the training data as a validation set and considered early stopping to avoid over-fitting. 

\begin{figure}[ht]
  \subfigure[]{\includegraphics[width=\linewidth]{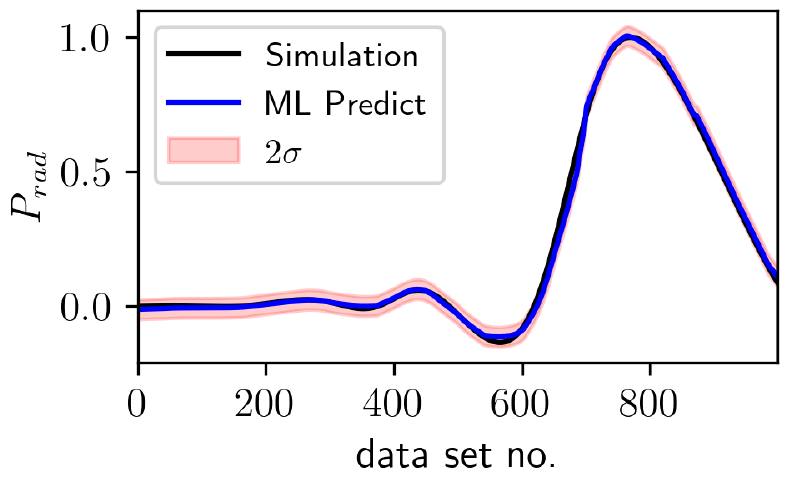}}
  \subfigure[]{\includegraphics[width=\linewidth]{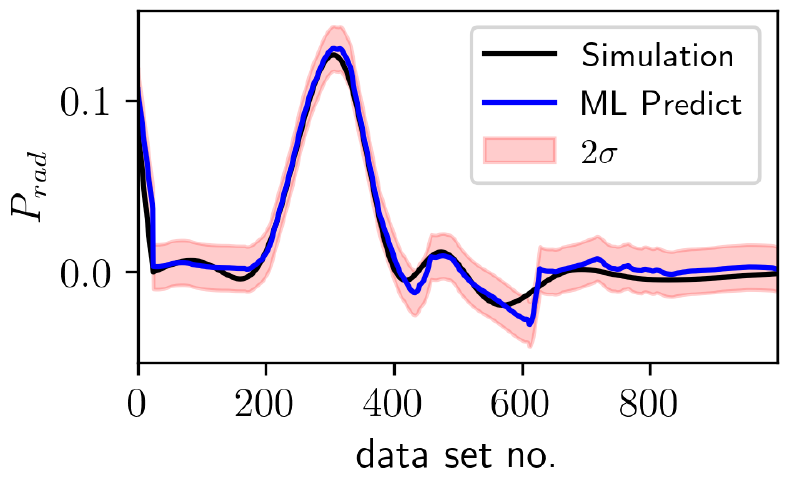}}
\caption{Normalized radiation pressure predicted by the network for different sets of known input data (source frequency, dielectric permittivity) along with true data from simulation. Radiation pressure is normalized with maximum value available in the training data set. Top (a) and bottom (b) panels represent the network performance of two randomly selected series of data sets.}
\label{fig:11}
\end{figure}

To test our DNN network, we randomized the input data set keeping the output data tethered and created two unique data set comprised of 1000 data points. A comparison between the simulated data and the ML predicted data is shown on \cref{fig:11}. The predicted values are found to lie within $2\sigma$ limit. 

The DNN source code has been made available under MIT License. User can find the relevant details by visiting Zenodo or GitHub\cite{Mishra_deepRadiation_2022} (\href{https://doi.org/10.5281/zenodo.6300930}{doi:10.5281/zenodo.6300930}).
\begin{figure}[ht]
{\includegraphics[height=\linewidth]{Fig9.pdf}}
\caption{A visual representation of our DNN model. (\href{https://raw.githubusercontent.com/Neural-Plasma/deepRadiation/main/figures/Fig9.png}{Click to zoom})}
\label{fig:9}
\end{figure}

\section{Conclusions}
Our simulation model confirms the existence of the surface wave by the evanescent nature of the field. The principal focus of the present study was to examine the properties of the surface wave for different source frequency and permittivity values. It has been observed that at particular cutoff frequency, the model supports the charge bunching indicating the existence of surface wave. Below that frequency the system does not show any bunching or surface wave propagation.  We have also noticed, with increasing frequency the bunching width ($\delta$) starts to decrease . The material permittivity has also a significant effect on the charge separation. Increasing permittivity for the surrounding material decreases the charge separation width. The presence of surface wave has some adverse effects on the electron dynamics. The energy of the electrons throughout the system has been studied using the reconstructed velocity distribution. Near the interface the electrons seem to follow the Kappa-Cairns distribution, whereas, in the bulk region of plasma (near to origin), they approach the Maxwellian distribution. However, the distribution was slightly deviated from Maxwellian, which we suspect due to small influx of higher energy particles towards the bulk region. The electron heating present in the system is very much different from the usual discharge process as it can be controlled by the supplied input frequency and the permittivity value. The presence of the surface wave field is responsible for the hot electron population in the near interface region. The parameters considered for the study also seem to affect the radiation pressure which constitutes the energy loss in the system. From the observations, we can suggest that the large permittivity material might not be a good choice for the surface wave study. The above observation demonstrate that the sustainment of the surface wave is configurable, hence can be controlled by physical parameters.

Deep Neural Network (DNN) has been introduced as a possible alternative to running PIC simulations for estimation of radiation pressure. For the present scenario, the degrees of freedom (DNN model) is two (source frequency and permittivity). The ideal would have been to include the length and radius of the system, which would increase the total number of runs multi-fold. With a moderate statistical noise for 1024 parametric variations ($\sim$ 20 hours per run), it took almost 42 days using HPC (parallel and sequential). Keeping the run-time and computational budget in mind, we tried to keep the total number of inputs (degree of freedom) limited to the case. In our opinion, the present way of using the PIC simulation as the data acquisition method for constructing a DNN model might be an impractical approach due to its computational cost. However, the open-source nature of the code and flexible input parameter space can allow people to expand the model to build on top of the existing datasets without rerunning the entire parameter space. 

\begin{acknowledgments}
The XOOPIC simulations were performed on ANTYA HPC Linux cluster at the Institute for Plasma Research (IPR). The authors would also like to thank the Computer Center staff of IPR.
\end{acknowledgments}

\section*{DATA AVAILABILITY}
The data that support the findings of this study are available from the corresponding author upon reasonable request.

\section*{References}
\bibliographystyle{unsrt}
\bibliography{bibliography}

\end{document}